# Writing and erasing skyrmions by single ultrafast laser pulses in monolayer Janus 2D magnets

Guangyao Miao[1,2,6,7], Yonglong Ga[3,7], Chang Liu[1], Pan Chen[1], Yichen Jin[2], Florian Kronast[4], Wenxin Cheng[1,5], Zhaoqing Ding[1,5], Kai Hu[1], Zongnan Zhang[1], Nikolai Severin[2], Chenxi Meng[1], Patil Shubhada[4], Sergio Valencia[4], Meng Meng[1], Qinlin Guo[1], Xiaoran Liu[1], Jiandi Zhang[1,5], Yangmu Li[1,5], Carlos-Andres Palma[1,2], Jürgen P. Rabe[2], Hongxin Yang*[3], Weihua Wang*[1], Jiandong Guo*[1,5]

[1]Beijing National Laboratory for Condensed Matter Physics and Institute of Physics, Chinese Academy of Sciences; Beijing, China
[2]Department of Physics and Center for the Science of Materials Berlin, Humboldt-Universität zu Berlin; Berlin, Germany.
[3]Center for Quantum Matter, School of Physics, Zhejiang University; Hangzhou, Zhejiang, China
[4]Helmholtz-Zentrum Berlin für Materialien und Energie; Berlin, Germany
[5]School of Physical Sciences, University of Chinese Academy of Sciences; Beijing, China
[6] Present address: II. Physikalisches Institut, Universität zu Köln; Köln, Germany
[7]These authors contributed equally to this work

*Corresponding authors. Email: hongxin.yang@zju.edu.cn (H.Y.);
weihuawang@iphy.ac.cn (W.W.);
jdguo@iphy.ac.cn (J.G.);

**Abstract:** Skyrmions in 2D magnets are promising candidates for nonvolatile, low-power, and high-density spintronic memories. However, their experimental realization at the 2D limit remains challenging, owing to the difficulty in engineering the required chiral magnetic interactions. Here, we report the creation and direct imaging of Néel-type skyrmions in Janus 2D chromium chalcogenides using synchrotron X-ray photoemission electron microscopy, and scanning nitrogen-vacancy magnetometry, which exhibit field-free stability, nonvolatility, and size tunability. First-principles calculations and micromagnetic simulations reveal that Janus-surface-induced inversion-symmetry breaking enhances the Dzyaloshinskii–Moriya interaction, providing the microscopic mechanism for skyrmion stabilization and tunability. We further achieve reversible skyrmion writing and erasing using a single ultrafast laser pulse in a magnetic field as low as 300 *Oe*, demonstrating the excellent manipulability of this 2D magnetic system. These results establish Janus engineering as a route to creating and manipulating nonvolatile skyrmions in atomically thin magnets, with implications for skyrmion-based low-power spintronic devices.

## Main

Magnetic skyrmions have attracted tremendous interest due to their topological stability, nanoscale size, and low-current dynamics, making them promising candidates for applications in racetrack memory, logic devices, neuromorphic computing, and quantum computing[1-4]. Microscopically, the formation of such exotic magnetic textures typically arises from the interplay among symmetric Heisenberg exchange, antisymmetric Dzyaloshinskii-Moriya interactions (DMI), magnetic anisotropy, and dipolar interactions in systems with broken inversion symmetry and strong spin-orbit coupling (SOC)[5]. The magnetic skyrmions were initially discovered in B20-type bulk magnets with non-centrosymmetric chiral crystal structures[6,7]. Later studies developed thin-film systems composed of *5d* heavy-metal and *3d* ferromagnetic layers, even at the monolayer (ML) limit, e.g., ML-Fe/Ir[8]. In these systems, the interfacial DMI, driven by designed structural symmetry breaking and strong SOC, plays a crucial role in creating, stabilizing, and manipulating skyrmions. In parallel with these developments, research on magnetic skyrmions in pure van der Waals (vdW) magnets has advanced rapidly since the discovery of long-range ferromagnetic ordering in the 2D limit[9,10]. Benefiting from their unparalleled tunability, 2D atomically thin magnets offer advantages for future device applications, as their reduced dimensionality makes skyrmion topology, stability, and dynamics more directly controllable via interfaces, gating, strain, and stacking engineering. Although magnetic skyrmions have been experimentally discovered in several emerging van der Waals (vdW) and non-vdW 2D magnetic materials[11-18], almost all experiments to date have focused on bulk crystals or thick films.

Creating and engineering magnetic skyrmions as the magnets approach the 2D unit-cell thickness poses fundamental challenges. First, thermal fluctuations, as dictated by the Mermin-Wagner theorem, suppress long-range magnetic order unless compensated by introducing magnetic anisotropy, while a large magnetic anisotropy is not favorable for creating skyrmions. Second, most intrinsic 2D magnets, including $Fe_3GeTe_2$, $CrI_3$, and $Cr_2Ge_2Te_6$, etc., possess centrosymmetric crystal structures, which inherently preclude the existence of DMI, the key ingredient for chiral spin textures. Third, methods for precisely engineering 2D magnets to balance magnetic interactions are limited. Theoretical work on already synthesized 2D magnets has proposed several feasible strategies, including applying an electric field, intercalation, geometric engineering, and constructing heterostructures[19-22]. Experimentally, noncollinear spin textures have recently been discovered in twisted-stacked 2D magnetic systems[23]. As an emerging family of materials, 2D Janus transition-metal compounds exhibit intrinsic structural symmetry breaking, endowing them with novel properties such as second-harmonic generation, spontaneous electrical polarization, and Rashba splitting[24,25]. When Janus structures meet magnets, the symmetry-breaking and interfacial DMI make them highly promising candidates for hosting magnetic skyrmions near the 2D limit, according to the theoretical predictions[26-29].

In this article, we report the creation and manipulation of magnetic skyrmions in a Janus CrTeSe monolayer on atomically thin $Cr_{1+\delta}Te_2$. The Néel-type skyrmions are directly imaged using surface-sensitive detection techniques, including X-ray photoemission electron microscopy (XPEEM) and scanning nitrogen-vacancy magnetometry within a diamond probe (scanning NV). We experimentally demonstrate the engineering of skyrmion size through thickness design in the structure. Furthermore, we achieve localized, fully reversible writing and erasing of single skyrmions using individual femtosecond laser pulses in a magnetic field as low as 300 *Oe*.

## Fabrication and characterization of Janus ML-CrTeSe

As shown in Fig. 1a, the Janus monolayer ML-CrTeSe is fabricated by surface substitution reactions under ultrahigh vacuum (UHV) conditions, following the previously reported recipe for surface Se-Te substitution in tellurium and tellurides[30,31]. The surface Se/Te substitution is triggered by depositing Se on the as-grown $CrTe_2$ layer and annealing the sample in UHV conditions. This process is repeated several cycles until the surface Te atoms are fully replaced by Se. Figure 1b-d compares the atom-resolved STM images of the same sample in different substitution stages. The bright and dim spots in Fig. 1c correspond to Te and Se atoms, respectively. The homogeneous brightness of atoms in Fig. 1d indicates the complete surface Se/Te substitution. Cross-sectional scanning transmission electron microscopy (STEM) measurements, including high-angle annular dark-field (HAADF) and integrated differential phase contrast (IDPC) images in Fig. 1e,f, depict the side view of Janus ML-CrTeSe. As shown in Fig. 1g, the intensity profile along the dashed line in the HAADF image (Fig. 1e) exhibits an asymmetry: the top Se-atom column exhibits lower intensity than the Te-atom column due to its smaller atomic number, consistent with recent STEM results on Janus TMCs[25,31,32]. This is in sharp contrast to the symmetric intensity profile of pure ML-$CrTe_2$ (Supplementary Fig. 1a,b).

Figure 1h presents the X-ray photoelectron spectroscopy (XPS) survey near the Cr *2p* and Te *3d* of Janus CrTeSe and $CrTe_2$. A similar feature supports the persistence of Cr-Te bonds in Janus CrTeSe. The Se *3d* binding energy peaks of Janus CrTeSe exhibit lower binding energy than elemental Se, indicating the existence of Cr-Se bonds (Supplementary Fig. 1c). The fitted area ratio of Cr *2p* to Te *3d* peaks in the Janus CrTeSe is nearly twice that in $CrTe_2$, as shown in Fig. 1h, which is consistent with the expected structure. Furthermore, compared to $CrTe_2$, the binding energy of Te *3d* in Janus CrTeSe is ~0.3 eV higher, indicating the loss of electrons from the bottom Te atoms in the Janus structure. This is due to the higher electronegativity of Se relative to Te. As shown in Fig. 1i, the $dI/dV$-$V$ and $I$-$V$ spectra of Janus CrTeSe exhibit a 0.24 eV asymmetric gap below the Fermi level, in contrast to the dip feature in $CrTe_2$, supporting the existence of a vertical electric dipole from surface Se pointing to bottom Te in the Janus structure.

**Néel-type magnetic skyrmions in Janus ML-CrTeSe/$Cr_{1+\delta}Te_2$**

Experimentally, the Janus CrTeSe monolayer grown on graphene/SiC exhibits antiferromagnetic ordering with a regulated magnetic anisotropy and a charge density wave[31]. Here, the creation and manipulation of magnetic skyrmions in a Janus CrTeSe monolayer on few-layer $Cr_{1+\delta}Te_2$ films (denoted as Janus ML-CrTeSe/$Cr_{1+\delta}Te_2$, as sketched in Fig. 2a) is experimentally realized (see Supplementary Fig. 2 for the representative morphology). Given that robust ferromagnetism in ultrathin $CrTe_2$ and Cr-intercalated $Cr_{1+\delta}Te_2$ films has been experimentally confirmed from bilayer thickness[33], the underlying few-layer $Cr_{1+\delta}Te_2$ provides the system with a ferromagnetic background and degree of freedom in controlling magnetization, while the Janus CrTeSe monolayer on top introduces symmetry-breaking and enhanced DMI.

A representative Janus ML-CrTeSe/$Cr_{1+\delta}Te_2$ sample (thickness~3.9 nm) exhibits a ferromagnetic behavior with a Curie temperature ($T_c$) ~195 K, as determined from temperature- and field-dependent magnetization measurements, as shown in Fig. 2b,c. Employing XPEEM based on synchrotron radiation, which is sensitive to the surface magnetic momentum, the magnetic structures in Janus ML-CrTeSe/$Cr_{1+\delta}Te_2$ are characterized (see Supplementary Fig. 3a for a sketch of the experimental setup). As shown in Fig. 2d, the X-ray magnetic circular dichroism (XMCD)-PEEM image acquired on a 6.0 nm thick Janus ML-CrTeSe/$Cr_{1+\delta}Te_2$ sample after directly cooling down to 49.2 K from room temperature at zero magnetic field (ZFC) reveals clear ferromagnetic domains with opposite out-of-plane magnetizing directions (blue and red area), along with several isolated small circular magnetic domains inside these ferromagnetic domains (denoted as *domain state*).

Using in situ ultrafast laser irradiation with/without a magnetic field, demagnetization and rapid-quench (laser fluence is set to 8.56 mJ/cm$^2$) field cooling (FC), as well as slow-annealing (laser fluence is set as a linear gradient decreased from 8.56 mJ/cm$^2$ to zero at a rate of 0.535 mJ/cm$^2$ per second) field cooling processes, can be achieved (see Supplementary Fig. 3b,c). After rapid-quench FC with a 300 *Oe* out-of-plane magnetic field, isolated circular magnetic domains in a ferromagnetic background are observed in the XMCD-PEEM image (measured at magnetic remanence after removing the external field), as shown in Fig. 2e (denoted as the *skyrmion state 1*). Using a slow-annealing FC process, the ferromagnetic background is stronger but with fewer isolated circular magnetic domains, as shown in Fig. 2f (denoted as *skyrmion state 2*) (see Supplementary Fig. 3d,e for data of opposite magnetic field directions). The difference between the states after rapid-quench FC and slow-annealing FC lies in the density of isolated circular magnetic domains and the homogeneity. These isolated circular magnetic domains are randomly distributed and similar in size, rather than being pinned to specific sites (the adsorbate in Supplementary Fig. 3f for image alignment). As shown in Fig. 2g, the intensity profiles along the line crossing the circular magnetic domains in the direction of the X-ray incidence obtained after ZFC and FC at 300 *Oe* exhibit similar full-width-at-half-maximum (FWHM) values (~80 nm), suggesting a common physical origin. Furthermore, deviations of the XMCD asymmetry intensity from a Gaussian fit are observed in the left and right sides of the peak. It is consistent with the characteristics of Néel-type skyrmions in the XMCD-PEEM measurement geometry and is confirmed by the scanning NV magnetometry measurements discussed below. The states after ZFC, rapid-quench FC, and slow-annealing FC remain unchanged when a 300 *Oe* external field is applied.

Scanning NV measurements are performed to reconstruct the stray-field map of an isolated circular magnetic domain in Janus ML-CrTeSe/$Cr_{1+\delta}Te_2$ after FC (Supplementary Fig. 4a for the sketch). Experimentally, the Janus ML-CrTeSe/$Cr_{1+\delta}Te_2$ sample used in M-T/M-H measurements (Fig. 2b,c ) was cooled from room temperature to 50 K while maintaining an out-of-plane magnetic field of 300 *Oe*. The large-scale iso-*B* scanning NV image measured in zero magnetic field after FC, indicating a magnetic state with some isolated circular magnetic domains (Supplementary Fig. 4b). The measured 2D map of the stray field near a circular magnetic domain, projected along the direction parallel to the NV axis, exhibits features similar to those observed in the XMCD-PEEM, as shown in Fig. 2h. Following the method proposed by Y. Dovzhenko et al.[34], the 2D maps of the stray field in the *X*, *Y*, and *Z* directions are reconstructed from the full-*B* scan data. The stray field shows a radial feature (Fig. 2i-k), consistent with the spin structure and stray field of a Néel-type magnetic skyrmion (Supplementary Fig. 4c,d). The skyrmions in Janus ML-CrTeSe/$Cr_{1+\delta}Te_2$ exhibit field-free stability and nonvolatility, as evidenced by XMCD-PEEM and scanning NV measurements performed in the absence of an external magnetic field. Compared to skyrmions generated by a large external magnetic field, this field-free stability enhances the tunability and accessibility of the skyrmions in the Janus CrTeSe monolayer. The ultrafast-laser-driven FC discussed above is, in fact, a skyrmion-writing process.

**Tuning skyrmions in Janus ML-CrTeSe/$Cr_{1+\delta}Te_2$**

The surface Janus structure in the studied system induces a significantly enhanced DMI, which is rare in atomically thin magnets. Figure 3a displays the schematic of layer-resolved DMI coefficient $d_k$ *vs* layer k of $Cr_{1+\delta}Te_2$ and Janus ML-CrTeSe/$Cr_{1+\delta}Te_2$ films for 2 layers (2L). Using the real-space spin-spiral method[22,35], we first calculate the layer-resolved DMI strengths $d_1$, $d_2$, and $d_3$ to be 1.57, 0.00, and -1.57 meV for $Cr_{1+\delta}Te_2$ as shown in Fig.3b. DMI arises from the spin sublattices of surface Cr atoms, which locally break inversion symmetry. But the total DMI ($d_{tot}$) sums to zero as the global inversion symmetry is preserved. Further analysis of the SOC energy distribution $\Delta E_{SOC}$ associated with the $d_{tot}$, as shown in Fig. 3c, reveals that the DMI between the Cr atoms in the surface layer k is predominantly linked to the Te elements with strong SOC, whereas the Cr elements contribute negligibly. In contrast, the Janus ML-CrTeSe/$Cr_{1+\delta}Te_2$ film breaks the global inversion symmetry. In this case, the DMI arises from the surface and the intercalated Cr atoms, with the extracted DMI parameters $d_1$, $d_2$, and $d_3$ being 1.32, 0.96, and -1.52 meV, respectively. A pronounced $\Delta E_{SOC}$ associated with the $d_{tot}$ emerges at the Te (Se) layers near the surface. These behaviors are consistent with the Fert-Lévy mechanism[36,37], which describes the DMI in a typical noncentrosymmetric three-site triplet (e.g., Cr-Te(Se)-Cr) as arising from spin-orbit scattering on the heavy element located between two magnetic sites. Similar DMI enhancement behaviors are also observed in thicker films, e.g., 4L (Supplementary Fig. 5).

Topological spin textures in magnetic materials result from the competition among exchange interactions ($J$), $d_k$, and effective magnetic anisotropy ($K$). Reported that $Cr_{1+\delta}Te_2$ exhibits strong thickness-dependent $J$ and $K$[18,38], thickness naturally serves as a direct tuning parameter for skyrmions in Janus ML-CrTeSe/$Cr_{1+\delta}Te_2$. Experimentally, it is found that the thinner the film is, the smaller the skyrmion size, as shown in Fig. 3d. When the thickness is reduced to 1.9 nm (2L-$Cr_{1+\delta}Te_2$ underlying the ML-CrTeSe), the diameter of the skyrmions measured by XPEEM decreased to ~60 nm. Although it is close to the XPEEM's real-space resolution limit, several skyrmions remain visible against the ferromagnetic background and noise.

To elucidate the microscopic origin of thickness-dependent skyrmion size, we perform first-principles calculations to extract the key magnetic parameters of $J$, $d_k$, and $K$ as a function of thickness[1,39]. As shown in Fig. 3e, the DFT strength near the Janus side ($d_{surf}$) increases with the thickness of the underlying $Cr_{1+\delta}Te_2$. Meanwhile, the effective anisotropy $K$ decreases, and the exchange constant $J$ increases with thickness (see Supplementary Fig. 6**)**. Using these DFT-obtained magnetic parameters, micromagnetic simulations further reveal that field-free skyrmions can be stabilized at zero temperature in Janus ML-CrTeSe/$Cr_{1+\delta}Te_2$ films, and the skyrmion size systematically decreases as the thickness is reduced (Fig. 3f). The thickness-dependent skyrmion size is thus attributed to the thickness-mediated competition among $J$, $d_k$, and $K$, in agreement with the experimental results. To quantify the competition, the dimensionless parameter $\kappa = (\frac{4}{\pi})^2 \frac{2JK}{d_k^2}$ that describe the relative strength of exchange and magnetic anisotropy against the DMI (a smaller $\kappa$ corresponds to stronger DMI and a tendency to form chiral spin textures) is introduced following Ref.[40, 41], where the deduced $\kappa$ is 7.29 (3L), 2.82 (4L), 1.67 (5L) and 1.18 (6L). For $\kappa > 1$, a typical isolated metastable skyrmion can be created. Generally, the thickness-dependent relative strengths of DMI, exchange interactions, and magnetic anisotropy endowed the Janus ML-CrTeSe/$Cr_{1+\delta}Te_2$ with an additional degree of freedom for engineering magnetic states compared to thick films and bulk crystals. Furthermore, the XMCD-PEEM images acquired at 49.2 K under laser-pulse-train irradiation at different fluences indicate an increase in skyrmion size but the same feature and, at higher fluences, even melting into the *domain state,* which is reproduced by the temperature-dependent simulations of skyrmions in Janus ML-CrTeSe/$Cr_{1+\delta}Te_2$ (Supplementary Fig. 7).

**Writing and erasing skyrmions in Janus ML-CrTeSe/$Cr_{1+\delta}Te_2$**

The magnetic skyrmions in Janus ML-CrTeSe/$Cr_{1+\delta}Te_2$ can be reversibly written/erased by designing irradiation of the ultrafast laser and magnetic field (Fig. 4a): Applying an ultrafast laser pulse trains (8.56 mJ/cm$^2$) in the absence of an external magnetic field, the skyrmions in the shown area of Janus ML-CrTeSe/$Cr_{1+\delta}Te_2$ are erased and tuned to *domain states* (from the left panel to the middle panel). This erasure process can also be achieved by using a strong, single, ultrafast laser pulse. Here, the writing of skyrmions in the same area by applying a single ultrafast laser pulse in conjunction with a low magnetic field is demonstrated (from the middle panel to the right):

Applying a 300 *Oe* magnetic field and a single laser pulse irradiation (21.4 mJ/cm$^2$), the skyrmions are written on this area. This indicates that a single ultrafast laser pulse is sufficient to transiently demagnetize the Janus ML-CrTeSe/$Cr_{1+\delta}Te_2$ films. Furthermore, the local writing of skyrmions is confined to the area of ultrafast laser irradiation (Fig. 4b): only the area under ultrafast laser irradiation is tuned to host skyrmions, while the other area remains in the *domain state*.

The mechanism underlying the writing/erasing of skyrmions using an ultrafast laser pulse and magnetic fields involves demagnetization and magnetization path selection, as illustrated in Fig. 4c. As discussed above, the ZFC brings the system to a binary ferromagnetic *domain state*, with a small amount of skyrmions with opposite polarity observed in these two ferromagnetic domains due to the existence of DMI. With an ultrafast laser train or a strong single pulse, the Janus ML-CrTeSe/$Cr_{1+\delta}Te_2$ is transiently demagnetized to a highly nonequilibrium state. If no external magnetic field is applied, the system will return to the *domain state* after the ultrafast laser is removed. A low magnetic field (300 *Oe*) induces rapid-quench FC or slow-annealing FC, depending on the laser mode used, as discussed above. Both field-cooling protocols generate skyrmions within a field-aligned ferromagnetic background. Rapid-quench FC provides only a short relaxation window, freezing stochastic nucleation and local inhomogeneities into a metastable, nonuniform texture, whereas slow-annealing FC offers a much longer effective annealing time, allowing the magnetic texture to reorganize into a more uniform skyrmion distribution. Cooling-path control enables kinetic engineering of skyrmion populations, offering a simple route to tailor their density, spatial uniformity, and metastability for functional spintronic devices. This result demonstrates the potential to use optical patterns and local nanoscale magnetic heads to realize skyrmion-based memory devices in Janus ML-CrTeSe/$Cr_{1+\delta}Te_2$.

**Conclusion**

In summary, the creation and manipulation of magnetic skyrmions in a 2D monolayer of Janus magnetic chromium chalcogenides is demonstrated experimentally and theoretically. The mechanism of the structural design is revealed by the Janus CrTeSe monolayer, which introduces strong DMI due to intrinsic structural symmetry breaking, thereby competing with Heisenberg exchange interactions and the magnetic anisotropy of the underlying $Cr_{1+\delta}Te_2$, stabilizing skyrmions near the 2D limit. By tuning the thickness of ferromagnetic $Cr_{1+\delta}Te_2$, the skyrmion size is tuned to ~60 nm by engineering magnetic interactions. A prototype demonstration of using a single ultrafast laser pulse in conjunction with a low magnetic field to write/erase magnetic skyrmions is reported here, demonstrating the facile tunability of skyrmions in 2D atomically thin magnets.

Our work highlights the importance and potential of tuning the states of matter in 2D magnets via surface-atom engineering. In the future, by varying the Janus surface chemistry, interfacial coupling, and the species/thickness of the underlying ferromagnetic layers to tailor the balance of magnetic interactions, it is possible to control the formation, size, density, and stability window of skyrmions, even at room temperature. The demonstrated writing and erasing of skyrmions by a single ultrafast laser pulse shows the possibility of manipulating a single skyrmion using an integrated nanoscale magnetic head with optical stimuli in 2D Janus-engineered magnets. Time-resolved experiments to resolve the nonequilibrium nucleation and annihilation pathways and the possible magnetoelectrical response of the Janus structure are also worth exploring.

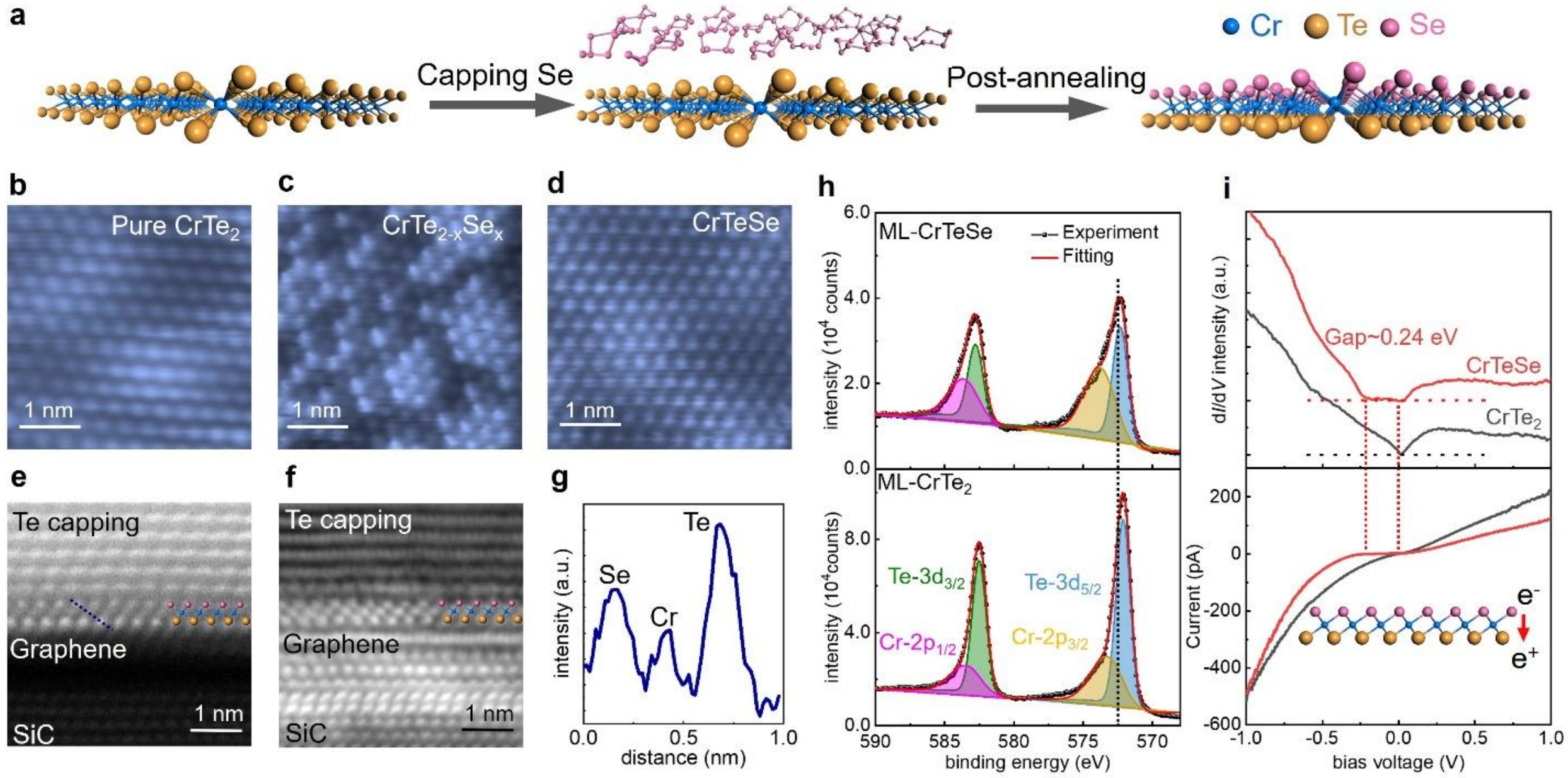

**Fig.1. The fabrication and characterization of Janus CrTeSe. a**, The sketch of the method for fabricating the Janus CrTeSe structure. **b-d**, The atom-resolved STM images of as-grown ML-$CrTe_2$ (**b**), ML-$CrTe_{2-x}Se_x$ with partial surface Se-Te substitutions (**c**), and Janus ML-CrTeSe (**d**). **e**, **f**, Cross-section high-angle annular dark-field (HAADF) (**e**) and integrated differential phase contrast (IDPC) (**f**) Scanning transmission electron microscopy (STEM) image of the Janus CrTeSe/Gr/SiC. The crystal structure is superposed on the image. **g** The intensity profile along the blue dashed line in (**e**). **h**, The XPS spectra near the binding energies of Cr *2p* and Te *3d* in ML-$CrTe_2$ and ML-CrTeSe. Black dots and red curves indicate the experimental data and fittings. To clearly show the shift in the Te $3d_{5/2}$ peak between the two samples, the CrTeSe peak is labeled with a black dashed line. **i**, The differential conductance (d$I$/d$V$-$V$) (upper panel) and conductance ($I$-$V$) (lower panel) spectra of $CrTe_2$ and CrTeSe. The d$I$/d$V$ spectrum of CrTeSe is shifted vertically for clarity. The black and red dotted lines indicate zero conductance. The inset illustrates the dipole moment.

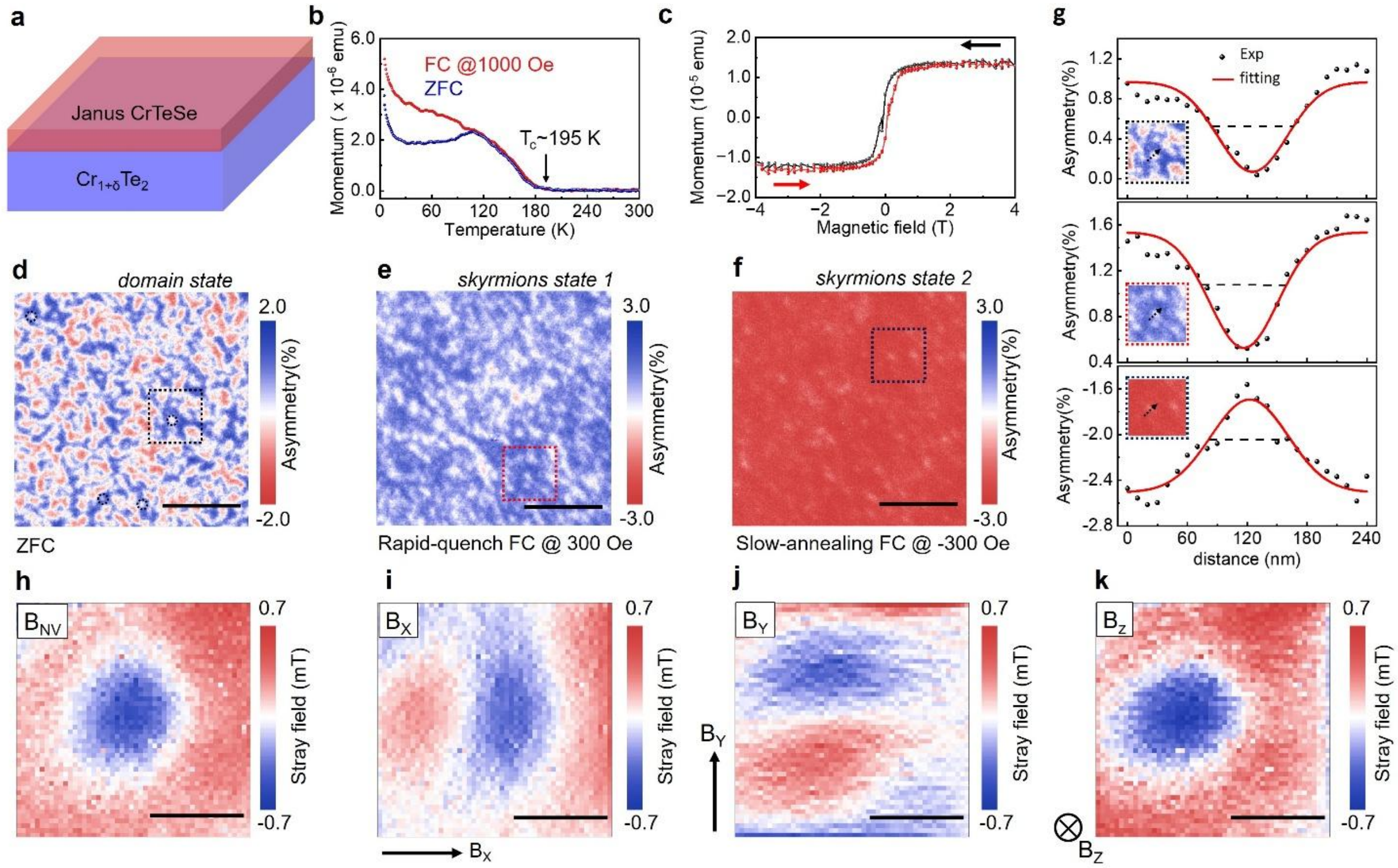

**Fig.2. Magnetic properties and Néel-type skyrmions in Janus ML-CrTeSe/$Cr_{1+\delta}Te_2$. a**, Sketch of the structure of Janus ML-CrTeSe/$Cr_{1+\delta}Te_2$. **b**, Temperature dependence of magnetization (*M*-*T*) curves for Janus ML-CrTeSe/$Cr_{1+\delta}Te_2$ films (3.9 nm, ~6L) under zero field cooling (blue), and field cooling by applying an out-of-plane magnetic field $H$ = 1000 *Oe* (red). **c**, Field-dependent magnetization (*M*-*H*) of Janus ML-CrTeSe/$Cr_{1+\delta}Te_2$ films under out-of-plane magnetic field measured at 50 K. **d-f**, The X-ray magnetic circular polarized dichroism photoemission electron microscopy (XMCD-PEEM) images (Cr-$L_3$ edge, 575.25 eV) of Janus ML-CrTeSe/$Cr_{1+\delta}Te_2$ films (6.0 nm, ~9L) measured after zero-field-cooling (ZFC) (**d**), rapid-quench FC with 300 *Oe* (**e**), and slow-annealing FC with -300 *Oe* (**f**). The scale bars of (**d**)-(**f**) are 1 μm. **g**, The intensity profiles along the arrows following the X-ray's direction across circular magnetic domains from (**d**)-(**f**). The raw data show left-right asymmetry in intensity compared to the Gaussian fitting. The insets show the zoomed-in XMCD-PEEM images. **h**, The 2D map of the stray field projection magnetic field parallel to the NV axis ($B_{//}$) of a magnetic skyrmion acquired by scanning NV measurements. The measurement was performed at a bias field of $B_{//ext}$ = 2 mT applied along the [111] diamond axis. **i-k**, The reconstructed components of the stray field along the $X$ (**i**), $Y$ (**j**), and $Z$ (**k**) directions according to full-$B$ NV measurements. The scale bars for (**h**)-(**k**) are all 200 nm.

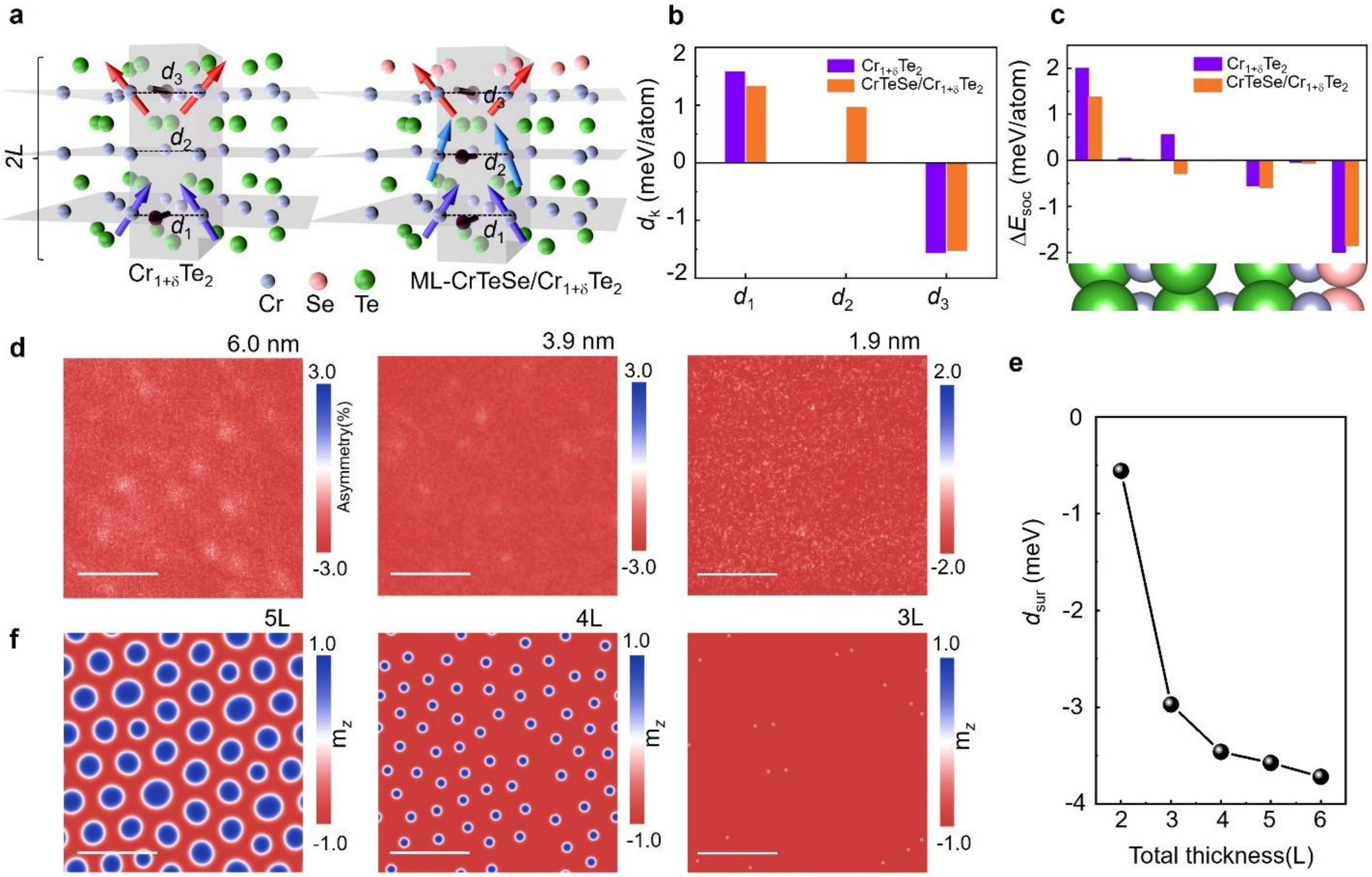

**Fig.3. DMI in Janus ML-$CrTeSe/Cr_{1+\delta}Te_2$ and thickness-engineered skyrmions' size**. **a**, Schematic of the spin structure of $Cr_{1+\delta}Te_2$ and $Cr_{1+\delta}Te_{2-x}Se_x$ structures under film thickness 2L. The blue, purple, and red arrows indicate the spins, while the black arrows indicate the DMI vectors. **b**, Layer resolved DMI coefficient $d_k$ vs layer k. **c**, SOC energy difference $\Delta E_{SOC}$ of $Cr_{1+\delta}Te_2$ and Janus ML-$CrTeSe/Cr_{1+\delta}Te_2$ under film thickness 2L. **d**, XMCD-PEEM images of 2D Janus ML-$CrTeSe/Cr_{1+\delta}Te_2$ after slow FC at -300 *Oe* with varying thickness. The thickness is controlled by varying the growth time while keeping the other growth parameters at 50 minutes, 30 minutes, and 15 minutes for 6.0 nm, 3.9 nm, and 1.9 nm thick samples, respectively. The scale bars are 500 nm. **e**, Calculated DMI parameters $d_{surf}$ as a function of film thickness. Here, $d_{surf}$ denotes the DMI strength associated with both the surface and the intercalated Cr atoms near the Janus-side surface. **f**, Micromagnetic simulation of skyrmions in Janus ML-$CrTeSe/Cr_{1+\delta}Te_2$ with varying thickness according to the DFT-calculated parameters. The scale bars are 500 nm.

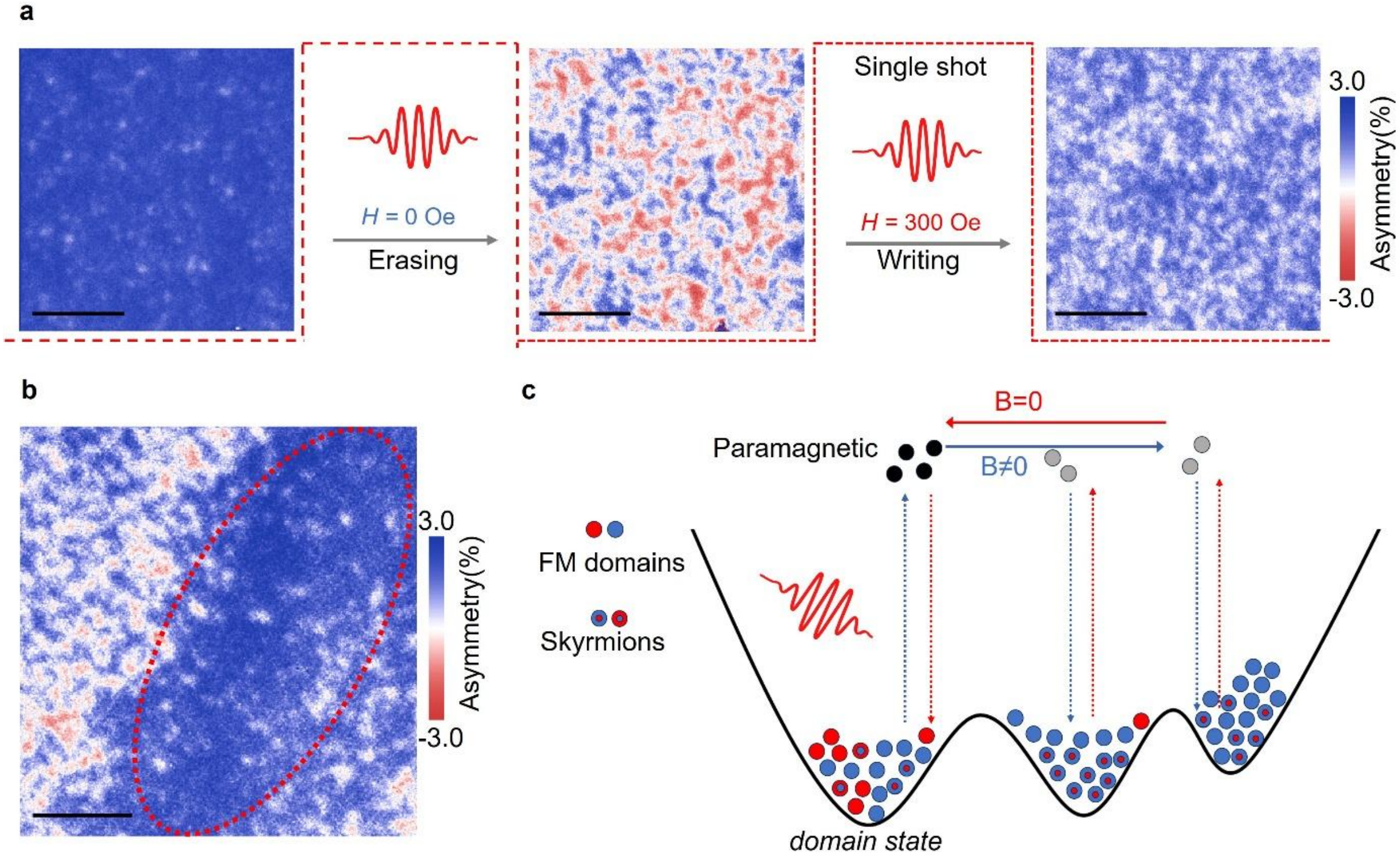

**Fig.4**. **Ultrafast-laser-driven writing/erasing skyrmions in Janus ML-CrTeSe/$Cr_{1+\delta}Te_2$**. **a**, Reversible erasing/writing skyrmions in Janus ML-CrTeSe/$Cr_{1+\delta}Te_2$ by a combination of magnetic field and laser irradiation. scale bar: 1 μm. **b**, The XMCD-PEEM image of an area contains both regions with and without irradiation of an ultrafast laser pulse train. The red-dashed ellipse roughly marks the area where laser irradiation occurred. Scale bar: 1 μm. **c**, The sketch of the mechanism using a laser to write/erase skyrmions in ultrathin Janus ML-CrTeSe/$Cr_{1+\delta}Te_2$ films. The red and blue solid balls are ferromagnetic domains with opposite magnetization directions. The blue (red) balls with a red (blue) dot inside indicate the skyrmions.

## Methods

**Sample preparation**: Janus ML-CrTeSe and Janus ML-CrTeSe/$Cr_{1+\delta}Te_2$ structures were fabricated in ultrahigh-vacuum (UHV) MBE-STM combined systems (Guo Cheng Inno) at base pressures below $3\times10^{-10}$ mbar. The 4H-SiC(0001) was degassed in the UHV chamber by annealing at 660 ℃ for 12h. The epitaxial few-layer graphene was prepared by high-temperature annealing of SiC substrates in an ultrahigh vacuum of $2\times10$ $mm^2$. High-quality Te (99.999+%, Alfa Aesar) was evaporated from a Knudsen cell. An electron-beam evaporator was used to evaporate Cr (99.95%, Trillion Metals Co. Ltd.). First, the ML-$CrTe_2$ and $Cr_{1+\delta}Te_2$ are grown on as-degassed 4H-SiC(0001) or graphitized 4H-SiC(0001), following the reported phase diagram[42]. Then, we construct the surface Janus structure in chromium telluride through surface Se-Te substitution, following a similar method to that in Ref.[30]. Step 1: Depositing Se on the grown chromium telluride. Step 2. Conducting post-annealing of the films in ultrahigh vacuum at ~150 ℃. Step 3. Repeat steps 1 and 2 to ensure that Se substitutes for almost all the surface Te atoms, forming the Janus structure in the topmost layer.

**Scanning tunneling microscopy/spectroscopy**: The as-grown samples were investigated by STM at 5 K (Unisoku) and 300 K (Guo Cheng Inno). The STM topographic images were acquired in constant-current mode with the bias voltage applied to the sample. To measure STS, the $dI/dV$ signals were acquired using a lock-in amplifier with a sinusoidal modulation of 1517 Hz at 10 mV.

**Atomic force microscopy**: AFM Imaging was performed with a Multimode instrument (Bruker Corporation) in scan-assist mode using ScanAsyst-Air cantilevers (Bruker Corporation). Images were processed with SPIP (Image Metrology). The images were line-flattened before extracting height profiles.

**Scanning transmission electron microscopy measurements**: STEM cross-sectional samples were prepared using focused-ion beam equipment (Thermofisher Scios 2) following a standard workflow. TEM characterization was performed in a double aberration-corrected JEOL Grand ARM 300. The HAADF-STEM image was acquired with a collection semi-angle of [54, 220] mrad and a convergence semi-angle of 22 mrad.

**X-ray photoelectron spectroscopy**: XPS experiments were conducted at room temperature using a Thermo Scientific ESCALAB 250X. The calibration of binding energy (BE) of the XPS conducted on Thermo Scientific ESCALAB 250 is calibrated concerning the pure bulk Au $4f_{7/2}$(BE=84 eV) and Cu $2p_{3/2}$ (BE=932.7 eV) lines. The BE is referenced to the Fermi level, calibrated using pure bulk Ni as Ef = 0 eV.

**X-ray photoemission electron microscopy**: XPEEM measurements were conducted at the UE49/PGMa beamline of the synchrotron radiation source BESSY II at the Helmholtz-Zentrum Berlin[43]. Real-space X-ray magnetic circular dichroism (XMCD)-PEEM measurements are conducted near the Cr $L_3$ edge (~575.25 eV). The XMCD asymmetry is defined as $(\sigma^{-}-\sigma^{+})/(\sigma^{-}+\sigma^{+})$, where $\sigma-$ and $\sigma+$ are the XAS signals at the maximum taken with left and right circularly polarized X-rays, respectively. The fixed angle of incidence of the incoming X-rays with respect to the sample surface was 16°. The magnetic contrast is proportional to the component of magnetization projected along the direction of X-ray propagation. An 800-nm titanium-sapphire laser with a 100-fs pulse duration and a 1.25-MHz repetition rate, focused on the same area as the X-ray beam, is used to demagnetize the sample. The laser spot was spatially overlapped with the X-ray beam and incident on the sample at a grazing angle of 16° with respect to the sample surface. The elliptical laser spot, measured on the sample surface, had FWHM full widths of approximately 11 μm and 30 μm along its minor and major axes, respectively. The nominal average fluence per pulse, calculated by dividing the pulse energy by the elliptical area defined by the corresponding $1/e^2$ beam radii, ranged from approximately 0.535 mJ/cm$^2$ to 21.4 mJ/cm$^2$, corresponding to average laser powers of 5 mW to 200 mW. Single-pulse and laser-pulse-train modes can be selected. The samples were capped with ~5 nm-thick amorphous Se protective layers and sealed in a rough vacuum during transfer. Post-annealing in UHV at 120 °C for 30 minutes to remove the capping layers before XPEEM experiments.

**Scanning nitrogen-vacancy magnetometers**: We perform magnetic field imaging using a scanning nitrogen-vacancy (NV) microscope, which operates from 4 K to 300 K and is integrated with a superconducting vector magnet[44]. A diamond tip containing a single NV center is mounted on an amplitude-modulation-based quartz tuning fork[45]. This assembly is combined with confocal optics that collect the photoluminescence (PL) from the NV center (with/without a microwave to resonate between the NV's sublevels) at every location. The NV center enables quantitative local magnetic field sensing through optically detected magnetic resonance (ODMR). The stray field component parallel to NV axis causes a net splitting between the $|m_s = 0\rangle$ and $|m_s = \pm1\rangle$ sublevels that follows $\Delta f = D \pm \gamma_e B_{\|}$, where $D \approx 2.87$ GHz represent zero-field splitting, and $\gamma_e = 28\ \mathrm{GHz \cdot T^{-1}}$ is the electronic spin gyromagnetic ratio[46-48]. This work utilizes two modes: the iso-$B$ mode for contour imaging of spin textures by fixing the microwave at the resonant frequency, and the full-$B$ mode for quantitative stray-field mapping by sweeping the microwave frequency to acquire full ODMR spectra[47]. During the experiment, the sample was first field-cooled from 300 to 50 K under a 30 mT external magnetic field applied along the Z-axis. After the field was reduced to zero, a 2 mT bias field was applied along the NV axis. In the full-$B$ mode, the stray-field distribution was obtained by determining the resonance frequency difference between the $|m_s = \pm1\rangle$ states. The NV center was positioned approximately 100 nm above the sample, with its orientation defined by a polar angle of about 54.7° and an azimuthal angle of 0° (i.e., aligned with the x-axis). The full vector stray field $B_x$, $B_y$, and $B_z$ were reconstructed using the spatially mapped distribution of $B_{\|}$[34,49].

**DFT calculations and micromagnetic simulations**: To study the magnetic properties of Janus ML-CrTeSe/$Cr_{1+\delta}Te_2$ films, we perform DFT calculations using the Vienna ab initio Simulation Package (VASP) [50]. The electron-core interaction is treated within the projected augmented wave (PAW) method[51-53], and the generalized gradient approximation (GGA) suggested by Perdew-Burke-Ernzerhof (PBE) is used to deal with the exchange-correlation effects[54]. To describe the strongly correlated effect between Cr 3*d* electrons well, the Hubbard $U$ value of 3.2 eV is chosen within the frame of the GGA+$U$ method[55,56]. A plane-wave cutoff energy of 520 eV and a Γ-centered 8×14×1 *k*-point are used for sampling the Brillouin zone. Structural relaxations are carried out until the total energy and Hellmann–Feynman forces converged to $1\times10^{-7}$ eV and 0.001 eV/Å, respectively. The method to obtain DMI parameters from the real-space spin spiral method is the same as described previously[22,35], in which a 1 × 4 × 1 supercell within Γ-centered 9 × 4 × 1 *k*-point sampling is used. To determine the ground states of the studied systems, four different magnetic configurations with a 1 × 2 × 1 supercell are considered for Janus ML-CrTeSe/$Cr_{1+\delta}Te_2$ films (see Supplementary Fig. 8). The calculations show that the FM configuration has the lowest energy, which is consistent with our experimental results. The effective anisotropy $K$ results from two competing contributions: SOC-induced magnetocrystalline anisotropy $K_{\mathrm{MCA}}$ and magnetostatic shape anisotropy $K_{\mathrm{MSA}}$ arising from dipole-dipole interactions. Micromagnetic simulations are subsequently conducted using the MuMax3 software package[57], with magnetic parameters obtained from first-principles calculations. In all simulations, a 1500×1500 nm$^2$ square with periodic boundary conditions was used to investigate real-space spin configurations. The details are provided in the Supplementary Text of Supplementary Information.

**Acknowledgment**

G.M. acknowledges the support from the IOP-Humboldt postdoc fellowship in physics. We thank the Helmholtz-Zentrum Berlin für Materialien und Energie for the allocation of synchrotron radiation beamtime (Nos. 242-12731, and 251-13159).

**Funding**: This work was supported by the National Key Research and Development Program of China (Grant No. 2024YFF0508503, No. 2022YFA1403400, Grants No. T2495212), NSFC (No. U23A20366, No. 12250710675, No. 12274439), Postdoctoral Fellowship Program of CPSF (No. GZC20232944, Grant No. GZC20252195). J.P.R. acknowledges the support of the Deutsche Forschungsgemeinschaft (DFG, German Research Foundation) Cluster of Excellence ‘Matters of Activity. Image Space Material’ (No. 390648296).

**Author contributions**: G.M., W.W., and J.G. conceived this project. G.M. led the research under the supervision of J.G., J.P.R., and W.W.. G.M., W.W., and Z. Z. carried out the MBE growth with the assistance of C.M.. G.M. and W.W., conducted the STM measurements. G.M., Y.J., F.K., and C.-A.P. performed the XPEEM measurements at BESSY II with assistance from P.S. and S.V. C.L., W.C., and Y.L. conducted the scanning NV measurements. Z.D., M.M., and X.L. contribute to the magnetization measurements. N.S. conducted the AFM measurements. K.H., P.C., and J.Z. designed and performed the STEM measurement. Y.G. and H.Y. designed and performed DFT calculations and micromagnetic simulations. Q.G. performed the XPS measurements. G.M., Y.G., W.W., H.Y., and J.G. wrote the manuscript with input from all other co-authors. All authors contributed to the discussion of the results.

**Competing interests:** The authors declare no competing interests.

**Data availability:** All data supporting the findings of this study are available in the main text and the Supplementary Information. Requests for additional data analysis should be directed to the corresponding authors

**Supplementary information**: Calculation details and more experimental data are included.

# Supplementary information for

**Writing and erasing skyrmions by single ultrafast laser pulses in monolayer Janus 2D magnets**

Guangyao Miao[1,2,6,7], Yonglong Ga[3,7], Chang Liu[1], Pan Chen[1], Yichen Jin[2], Florian Kronast[4], Wenxin Cheng[1,5], Zhaoqing Ding[1,5], Kai Hu[1], Zongnan Zhang[1], Nikolai Severin[2], Chenxi Meng[1], Patil Shubhada[4], Sergio Valencia[4], Meng Meng[1], Qinlin Guo[1], Xiaoran Liu[1], Jiandi Zhang[1,5], Yangmu Li[1,5], Carlos-Andres Palma[1,2], Jürgen P. Rabe[2], Hongxin Yang*[3], Weihua Wang*[1], Jiandong Guo*[1,5]

[1]Beijing National Laboratory for Condensed Matter Physics and Institute of Physics, Chinese Academy of Sciences; Beijing, China

[2]Department of Physics and Center for the Science of Materials Berlin, Humboldt-Universität zu Berlin; Berlin, Germany.

[3]Center for Quantum Matter, School of Physics, Zhejiang University; Hangzhou, Zhejiang, China

[4]Helmholtz-Zentrum Berlin für Materialien und Energie; Berlin, Germany

[5]School of Physical Sciences, University of Chinese Academy of Sciences; Beijing, China

[6]II. Physikalisches Institut, Universität zu Köln; Köln, Germany

[7]These authors contributed equally to this work

*Corresponding authors. Email: hongxin.yang@zju.edu.cn (H.Y.);
weihuawang@iphy.ac.cn (W.W.);
jdguo@iphy.ac.cn (J.G.);

## The file includes:

Supplementary Text
Figs. S1 to S10
References

## Supplementary Text

First, we investigate the bulk $Cr_{1+\delta}Te_2$. The calculated lattice constants a and $b$ are 7.04 Å and 4.09 Å. The magnetic ground state is determined by comparing four collinear spin configurations [ferromagnetic (FM) and A-, C-, and G-type antiferromagnetic (AFM)] in Fig. S8a-d, among which the FM state is found to be the energetically most favorable. Next, the thickness-dependent evolution of lattice constants in $Cr_{1+\delta}Te_2$ and Janus ML-CrTeSe/$Cr_{1+\delta}Te_2$ films is investigated, as shown in Fig. S9. One can see that the lattice constant of $Cr_{1+\delta}Te_2$ films increases and gradually approaches the bulk lattice constant with increasing layer thickness. This is consistent with the observation in the experiment. In particular, the substitution of top-layer Te atoms by Se in a $Cr_{1+\delta}Te_2$ thin film yields a Janus ML-CrTeSe/$Cr_{1+\delta}Te_2$ structure, thereby introducing an intrinsic spatial inversion asymmetry. A similar thickness-dependent increase in the lattice constant is also observed in Janus ML-CrTeSe/$Cr_{1+\delta}Te_2$ films. Notably, the lattice constants of the Janus ML-CrTeSe/$Cr_{1+\delta}Te_2$ films are systematically slightly smaller than those of their $Cr_{1+\delta}Te_2$ counterparts at the same film thickness. This trend is reminiscent of experimental observations in $MoS_2$ and the Janus MoSSe system[1].

Otherwise, to determine the magnetic ground states of $Cr_{1+\delta}Te_2$ and Janus ML-CrTeSe/$Cr_{1+\delta}Te_2$ films, we also consider four possible collinear magnetic configurations in Fig. S8a-d. Calculated results in Fig. S8e show that the FM state always keeps the ground states, which is consistent with experimental results.

To obtain a clear understanding of the distribution and associated SOC energy of DMI, we use the real-space spin spiral method [2,3]. This method has been successfully adopted for both bulk materials and interfaces[4,5]. In detail, a 1×4×1 supercell with 9×4×1 ***k***-mesh is adopted for obtaining the DMI parameters. We calculate the self-consistent energies of supercell $E_{cw}$ with clockwise {(0,0,$S$); (0, $S$, 0); (0,0,-$S$); (0, -$S$, 0)} and $E_{acw}$ with anticlockwise {(0,0,$S$); (0, -$S$, 0); (0,0,-$S$); (0, $S$, 0)} spin configurations. The total DMI strength $d_{tot}$ (Fig. S10a,b) and layer-resolved DMI strength $d_k$ (Fig. S10c,d) are solved by using opposite chirality spin configurations, respectively. Then, the DMIs stemming from the surface and the intercalated Cr atoms can be written as: $d_k = (E_{cw} - E_{acw})/24$ and $d_{k-1} = (E_{cw} - E_{acw})/8$, respectively. Here, a positive DMI value favors a spin configuration of anticlockwise chirality, and a negative one represents clockwise chirality.

We define the total energy difference between in-plane and out-of-plane magnetization directions as the magnetocrystalline anisotropy $K_{MCA}$. Otherwise, magnetic shape anisotropy $K_{MSA}$ arising from the dipole–dipole interactions is also considered as the contribution to $K$, and $K_{\mathrm{MSA}}$ can be calculated by $-\frac{1}{2}\left\{\frac{\mu_0}{4\pi V_{u.c.}}\right\}\sum_{i,j=1}^{N}\frac{(\boldsymbol{m}_i\cdot\boldsymbol{m}_j)r_{ij}^5-3(\boldsymbol{m}_i\cdot\boldsymbol{r}_{ij})(\boldsymbol{m}_j\cdot\boldsymbol{r}_{ij})}{r_{ij}^5}$, where $\boldsymbol{m}_i$ and $\boldsymbol{m}_j$ represent the unit vector of magnetization at position $\boldsymbol{r}_i$ and $\boldsymbol{r}_j$, $\boldsymbol{r}_{ij}$ is the unit vector between site $i$ and $j$. In calculations, we choose a $600 \times 600 \times 1$ supercell to obtain the dipole-dipole interaction between two magnetic atoms Cr. Then, $K$ can be expressed as $K = K_{\mathrm{MCA}} + K_{\mathrm{MSA}}$. Accordingly, $K > 0$ ($K < 0$) indicates the perpendicular magnetic anisotropy (in-plane magnetic anisotropy).

For obtaining the in-plane nearest-neighboring exchange parameters $J$, we compute the total energy differences of FM and G-type AFM states of a 1×2×1 supercell. The $J$ can be determined by mapping the total energy of different magnetic configurations based on following spin Hamiltonian: $H = J\sum_{i,j}\boldsymbol{S}_i\cdot\boldsymbol{S}_j + E_0$. Taking the structure with $N$ = 1 as an example, the total energy of different magnetic configurations can be calculated by following equations: $J = (E_{(\mathrm{G\text{-}type\ AFM})} - E_{\mathrm{FM}})/36$. Otherwise, we adopt the sign convention that $J > 0$ ($J < 0$) indicates the FM (AFM) coupling.

We perform the MUMAX3 package[6], to conduct spin dynamics by solving the Landau-Lifshitz-Gilbert (LLG) equation:

$$\frac{\partial \boldsymbol{m}}{\partial t} = -\gamma \boldsymbol{m} \times \boldsymbol{H_{eff}} + \alpha\left(\boldsymbol{m} \times \frac{\partial \boldsymbol{m}}{\partial t}\right),$$

where $\gamma$ denotes the gyromagnetic ratio, $\alpha$ is the damping constant, and $\boldsymbol{m}$ represents the normalized magnetization vector. $\boldsymbol{H_{eff}}$ is the effective field and can be obtained by following equation: $\boldsymbol{H_{eff}} = -\frac{1}{\mu_0\mu_s}\frac{\partial H}{\partial \boldsymbol{m}}$, where $H$ represents the total spin energy of studied system and can be expressed as: $H = H_{\mathrm{exc}} + H_{\mathrm{DMI}} + H_{\mathrm{ani}} + H_{\mathrm{ext}}$. Respectively, these energy terms are exchange energy $H_{\mathrm{exc}}$, DMI energy $H_{\mathrm{DMI}}$, anisotropy energy $H_{\mathrm{ani}}$, and external magnetic field $H_{\mathrm{ext}}$.

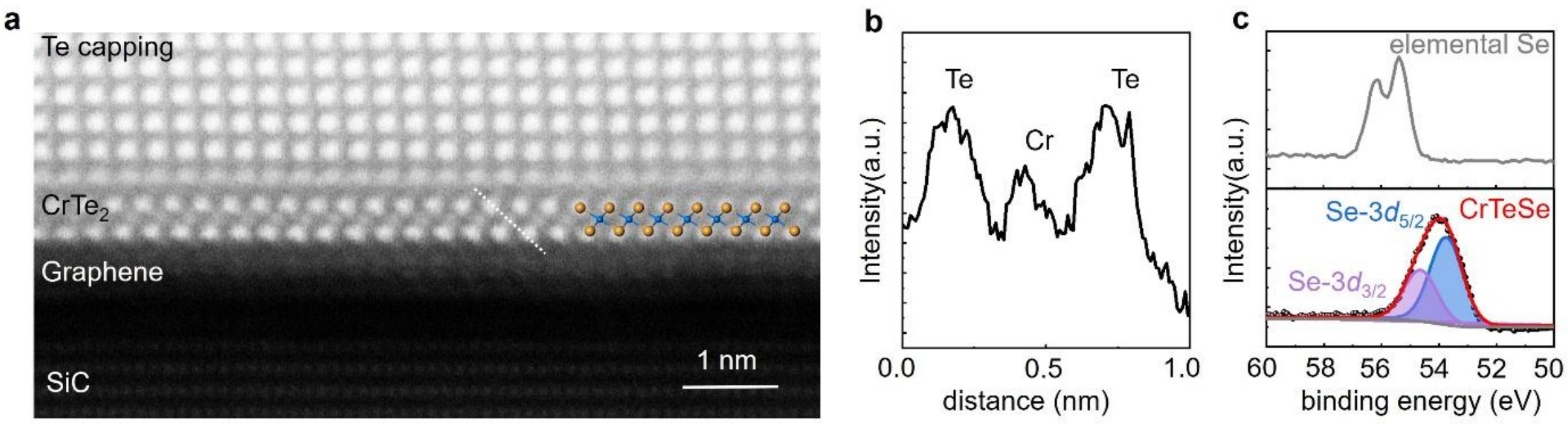

**Fig.S1: Scanning transmission electron microscopy (STEM) image of $CrTe_2$ and XPS results of Se 3*d* in Janus ML-CrTeSe and elemental Se**. **a**, High-angle annular dark-field (HAADF)-STEM image of $CrTe_2$. **b**, The intensity profile along the white dashed line in (**a**). **c**, XPS spectra of the Se 3*d* core level in Janus CrTeSe (lower panel) and elemental Se (upper panel). The black dots represent the experimental data, while the purple and blue lines correspond to the fitted Se $3d_{3/2}$ and Se $3d_{5/2}$ components, respectively. The red line represents the sum of the fitted components.

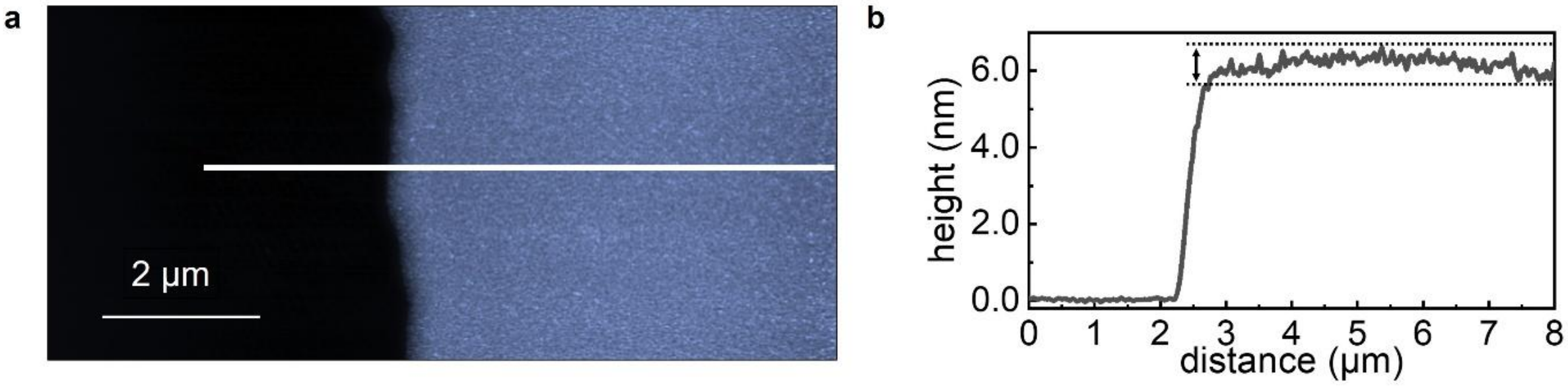

**Fig.S2: Morphology of Janus ML-CrTeSe/$Cr_{1+\delta}Te_2$**. **a**, Representative large-scale atomic force microscopy image of Janus ML-CrTeSe/$Cr_{1+\delta}Te_2$ after measuring XMCD-PEEM. **b**, Height profile along the white line indicated in (**a**).

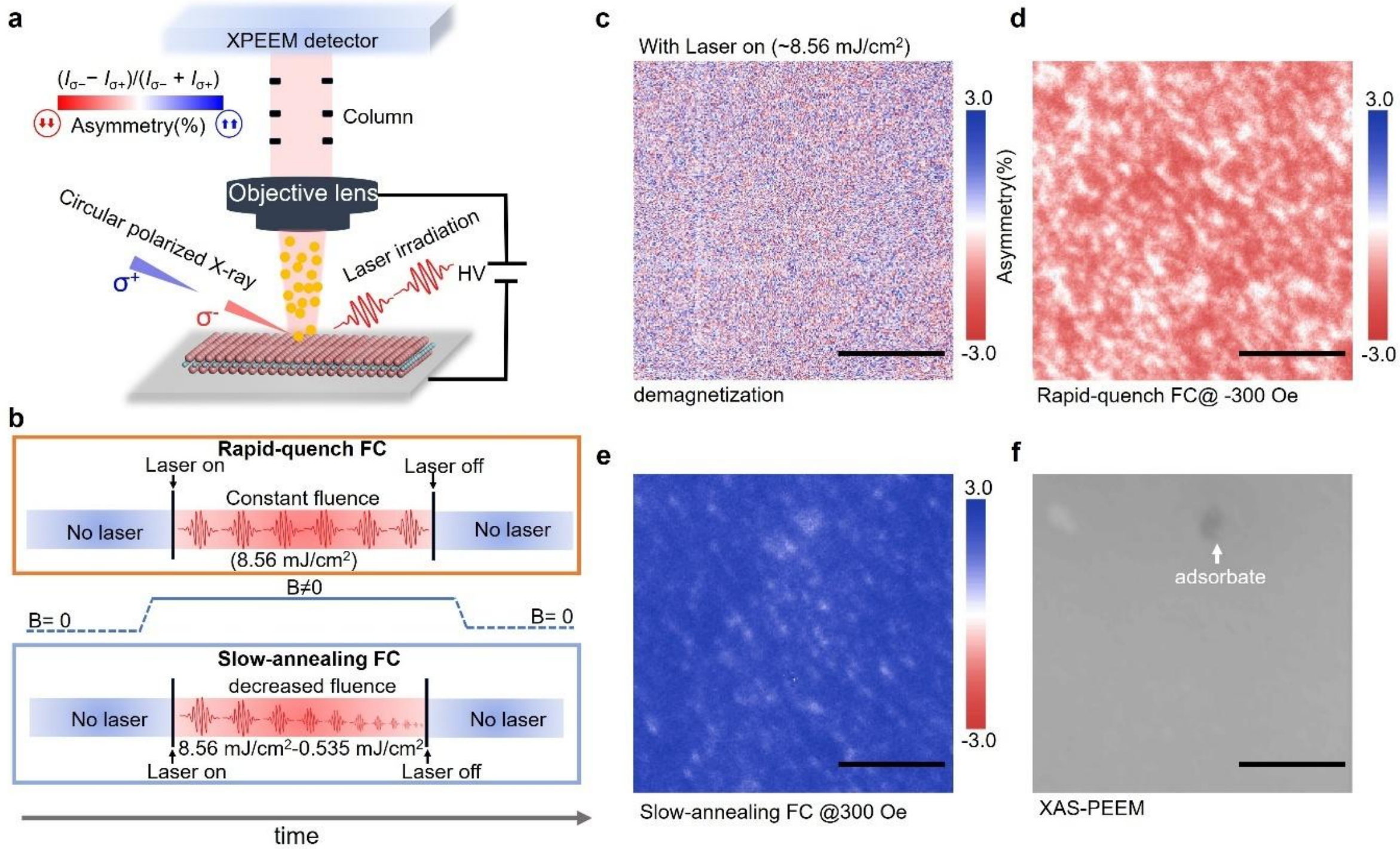

**Fig.S3: Sketch of XPEEM geometry, rapid-quench FC, slow-annealing FC, and more XPEEM data of Janus ML-CrTeSe/$Cr_{1+\delta}Te_2$.** **a**, The sketch of manipulating and detecting magnetic structure using XPEEM and an in-situ laser irradiation. **b**, Sketch of the rapid-quench field cooling (FC, upper panel) and slow-annealing FC (lower panel) processes in the XPEEM measurements. **c**, XMCD-PEEM image of the sample measured with ~8.56 mJ/cm$^2$ laser irradiation. **d** and **e**, The XMCD-PEEM images measured in the same area as Fig. 2**d-f** in the main text, after fast FC at -300 Oe (d) and slow FC at 300 *Oe* (**e**). **f**, XAS-PEEM image at a photon energy of 575.25 eV measured at the same area as **Fig. 2d-f** in the main text. The scale bars of (**c**)-(**f**) are all 1 μm

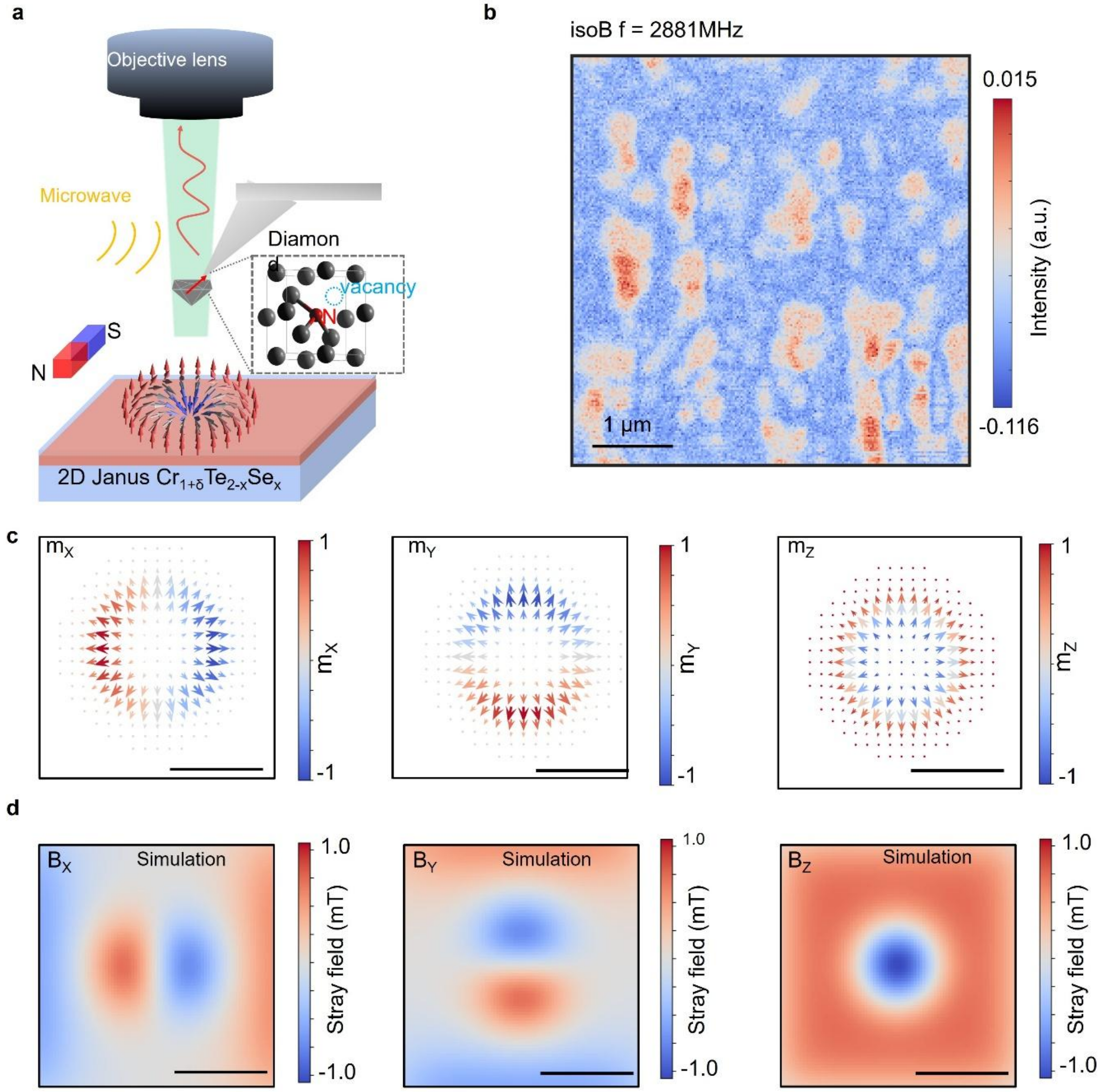

**Fig.S4: Scanning NV measurements and the spin structure of skyrmions in Janus ML-CrTeSe/$Cr_{1+\delta}Te_2$.** **a**, The sketch of the scanning NV measurements and the spin structure of a Néel-type skyrmion. **b**, The iso-*B* scan of Janus ML-CrTeSe/$Cr_{1+\delta}Te_2$ films (~3.9 nm, 6 L) using an NV magnetometer measured after cooling the sample from 300 K to 50 K with a 300 *Oe* out-of-plane magnetic field applied. Measured without a magnetic field at 50 K. **c**, The normalized magnetization of a Néel-type skyrmion along the *X*, *Y*, and *Z* directions. **d**, The simulated components of the stray field along the *X*, *Y*, and *Z* directions of a Néel-type skyrmion. The scale bars of (**c**) and (**d**) are 200 nm.

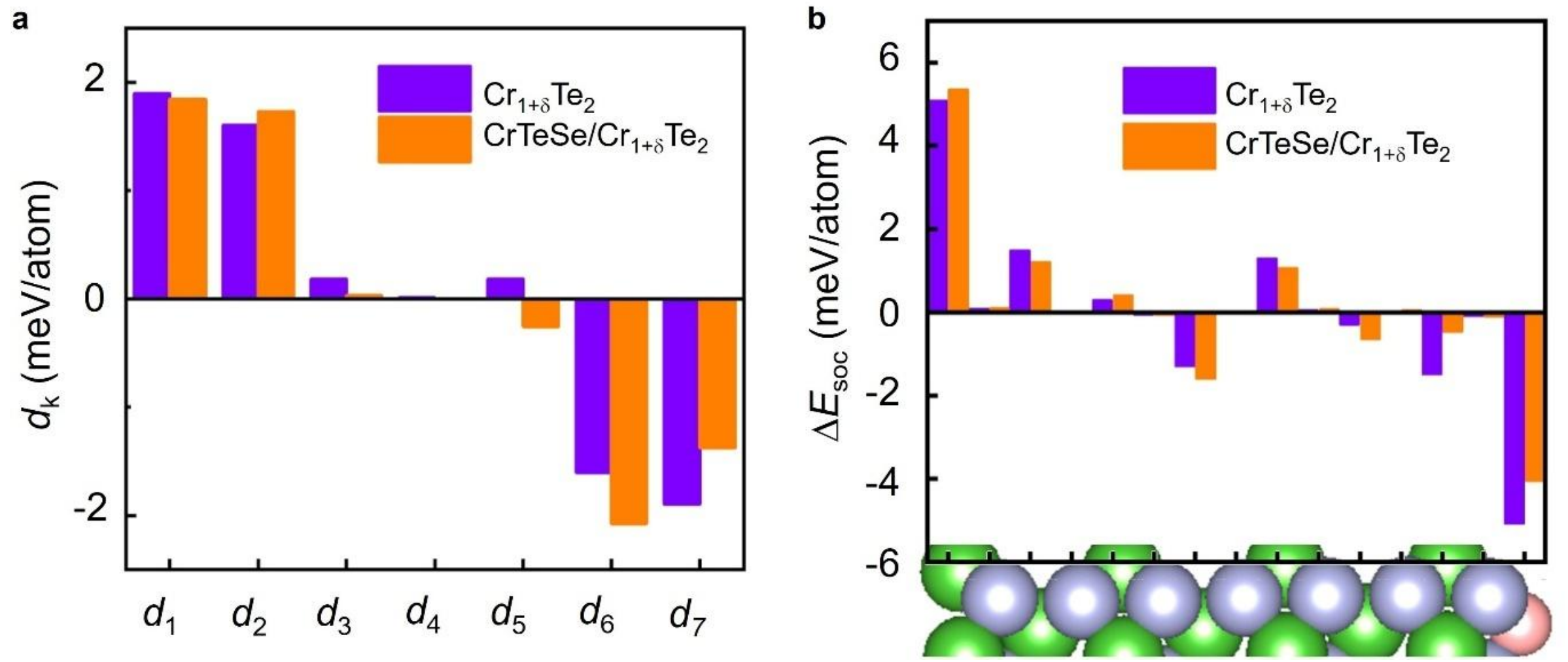

**Fig.S5: layer-resolved DMI and SOC in few layers $Cr_{1+\delta}Te_2$ and Janus ML-CrTeSe/$Cr_{1+\delta}Te_2$.** **a** and **b**, Layer-resolved DMI coefficient $d_k$ vs layer k (**a**), and SOC energy difference $\Delta E_{SOC}$ (**b**) of $Cr_{1+\delta}Te_2$ and Janus ML-CrTeSe/$Cr_{1+\delta}Te_2$ structures under film thickness of 4L.

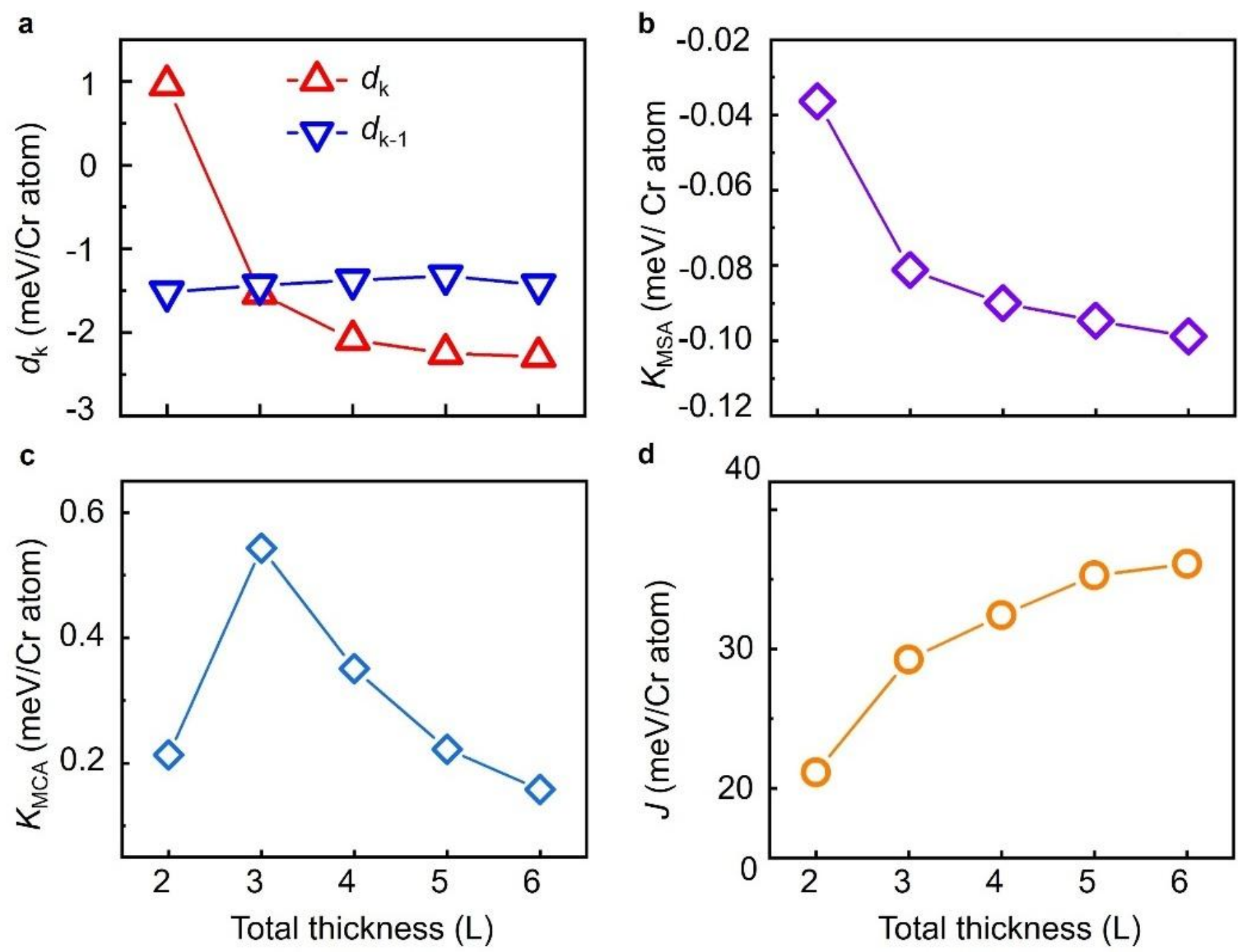

**Fig.S6: The calculated layer-resolved magnetic parameters of Janus ML-CrTeSe/$Cr_{1+\delta}Te_2$.** Layer-dependence of $d_k$ and $d_{k-1}$ near the Janus-side surface (**a**), magnetostatic shape anisotropy $K_{MSA}$ (**b**), magnetocrystalline anisotropy $K_{MCA}$ (**c**), and in-plane next-neighbor exchange parameter $J$ (**d**).

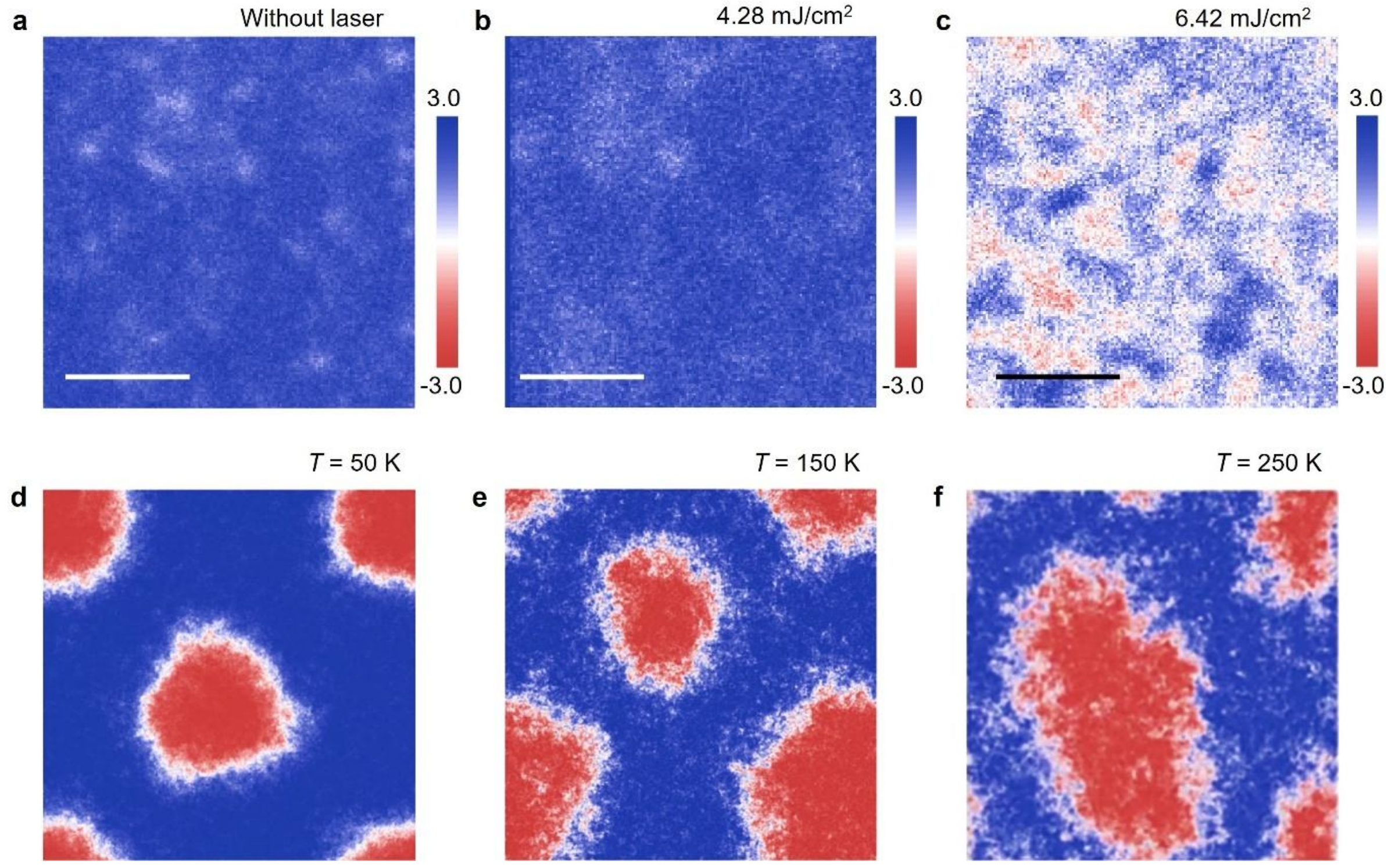

**Fig.S7: Temperature dependence of the skyrmions in Janus ML-CrTeSe/$Cr_{1+\delta}Te_2$. a-c,** XMCD-PEEM images of Janus ML-CrTeSe/$Cr_{1+\delta}Te_2$ measured with continuous ultrafast laser irradiation after slow FC at -300 *Oe*, without laser irradiation(**a**), with fluence ~4.28 mJ/cm$^2$ laser irradiation (**b**), with fluence ~6.42 mJ/cm$^2$ (**c**) laser irradiation. Measured at 49.2 K. The scale bars are 1 μm. **d-f** Micromagnetic simulation of skyrmions at 50 K (**d**), 150 K (**e**), and 250 K (**f**). The image sizes are 300×300 nm$^2$.

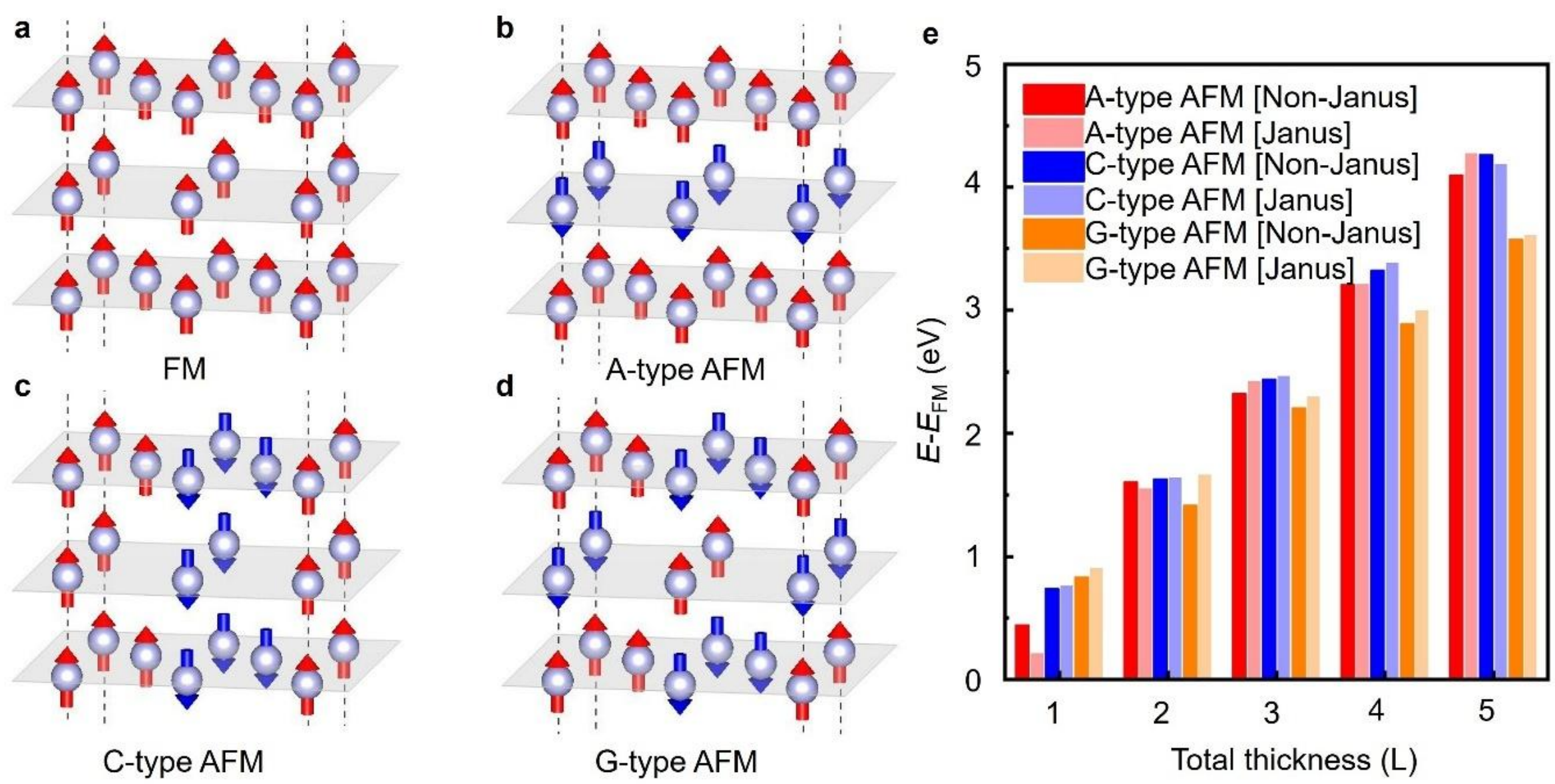

**Fig.S8**: **Calculated magnetic configuration of $Cr_{1+\delta}Te_2$ and Janus ML-CrTeSe/$Cr_{1+\delta}Te_2$. a-d**, Four different spin configurations, FM (**a**), A-type AFM (**b**), C-type AFM (**c**), and G-type AFM (**d**) of a 1 × 2 × 1 supercell are considered in our work. **e**, The energy differences relative to FM for different magnetic configurations are shown for the studied $Cr_{1+\delta}Te_2$ and Janus ML-CrTeSe/$Cr_{1+\delta}Te_2$ films.

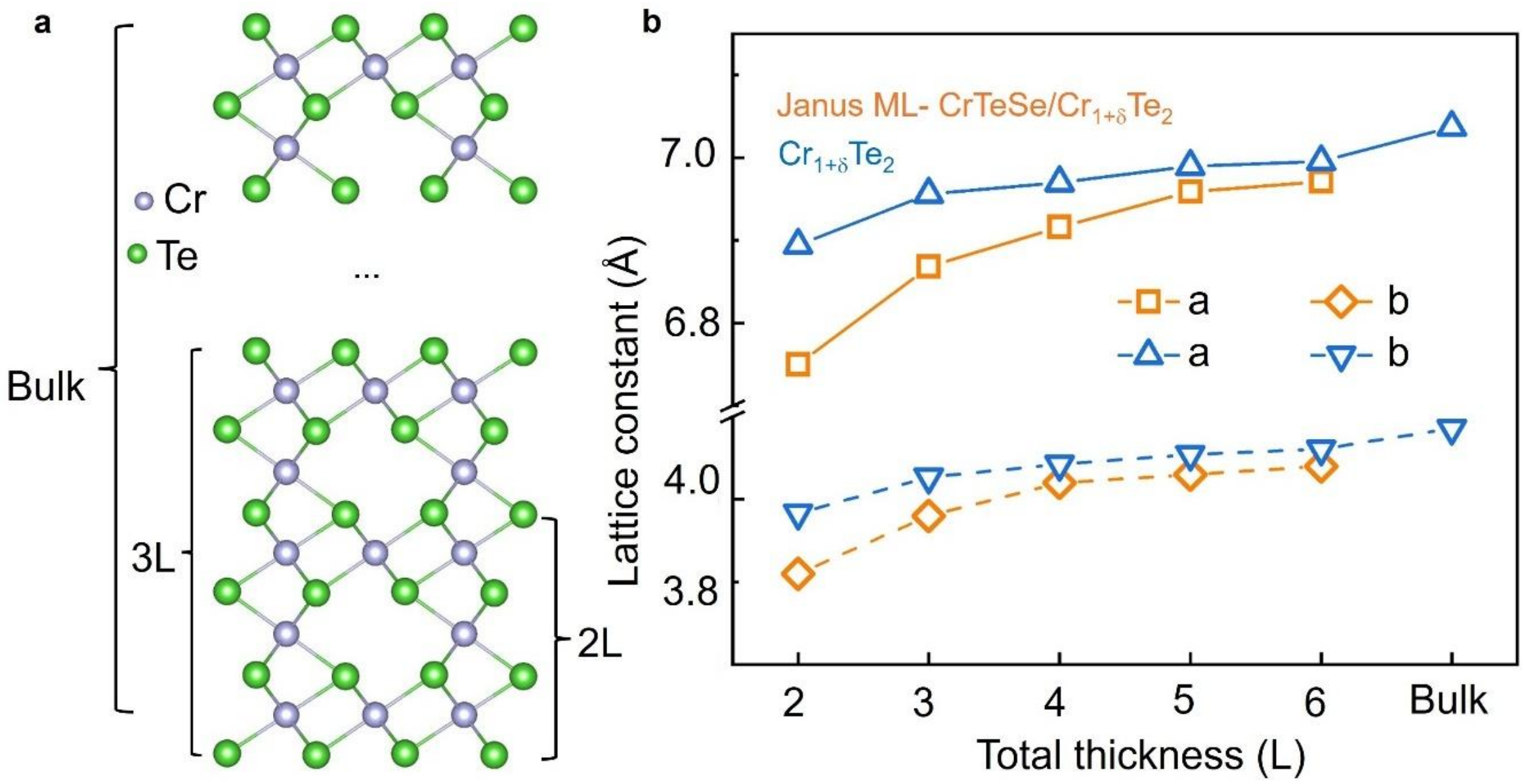

**Fig.S9: The definition of the thickness (a), and the thickness-dependence of lattice parameters (b) of the optimized $Cr_{1+\delta}Te_2$ and Janus ML-CrTeSe/$Cr_{1+\delta}Te_2$ structures**.

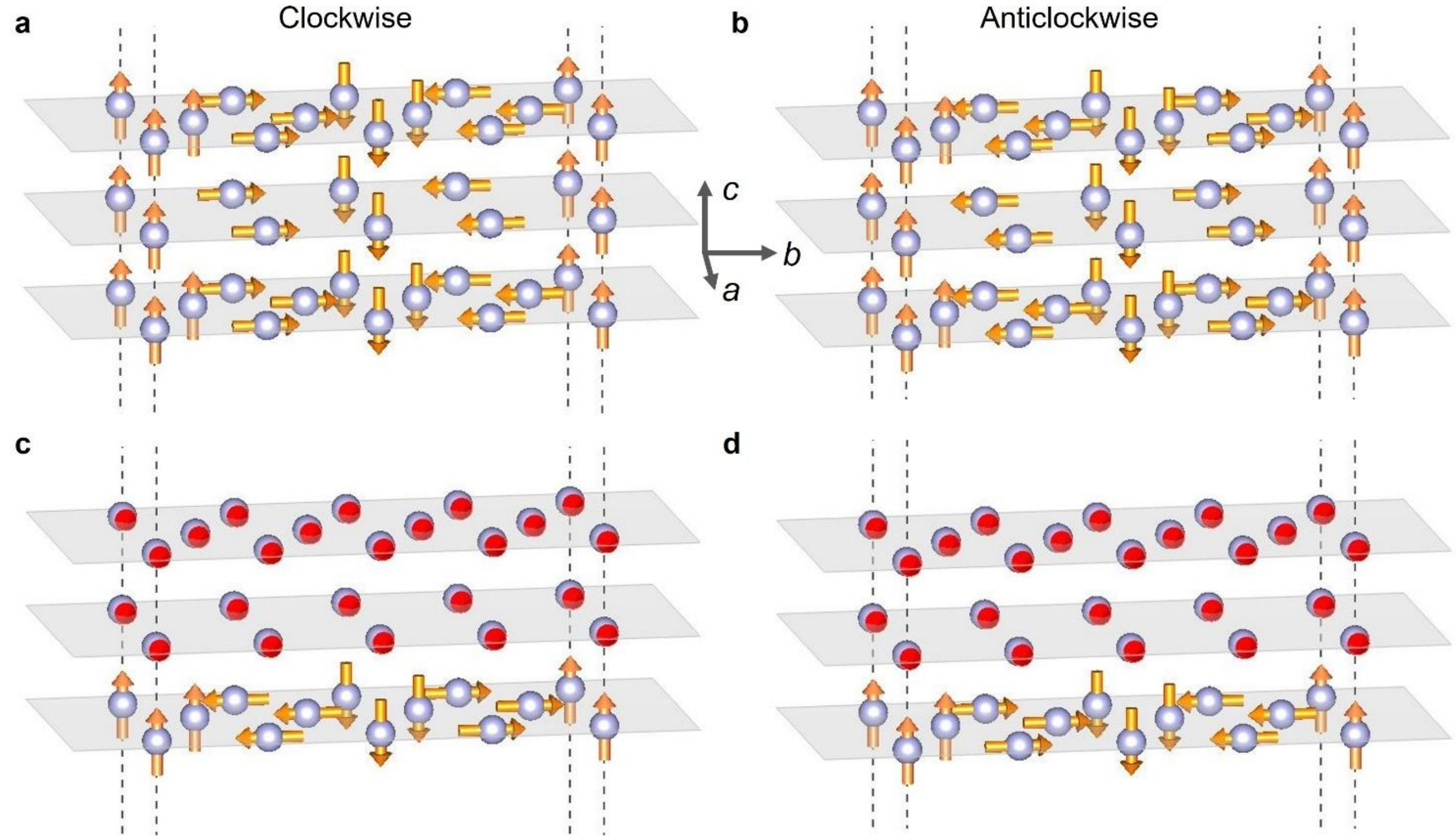

**Fig.S10**: **Clockwise (CW) and anticlockwise (ACW) spin configurations adopted to compute the DMI parameter of Janus ML-CrTeSe/$Cr_{1+\delta}Te_2$ films**. The total DMI $d_{tot}$, and layer-resolved $d_k$ can be extracted from the spin configurations in (**a**), (**b**), (**c**), and (**d**), respectively.